\documentclass[twocolumn,showpacs,preprintnumbers,amsmath,amssymb,preprint]{revtex4}
\usepackage{graphicx}% Include figure files
\usepackage{dcolumn}% Align table columns on decimal point
\usepackage{bm}% bold math
\usepackage{amssymb}
\usepackage{amsmath}
\begin{document}
\title{Sr$_2$IrO$_4$ magnetic phase diagram, from resistivity}

\author{L. Fruchter, G. Collin, D. Colson$^+$ and V. Brouet}%
\affiliation{Laboratoire de Physique des Solides, C.N.R.S. UMR 8502, Universit\'{e} Paris-Sud, 91405 Orsay, France}
\affiliation{$^+$Service de Physique de l'Etat Condens\'{e}, CEA-Saclay, 91191 Gif-sur-Yvette, France}
\date{Received: date / Revised version: date}
% The correct dates will be entered by Springer

\begin{abstract}{We show that, contrary to previous belief, the transition to the antiferromagnetic state of Sr$_2$IrO$_4$ in zero magnetic field does show up in the transverse resistivity. We attribute this to a change in transverse integrals associated to the magnetic ordering, which is evaluated considering hopping of the localized charge. The evolution of the resistivity anomaly associated to the magnetic transition under applied magnetic field is studied. It tracks the magnetic phase diagram, allowing to identify three different lines, notably the spin-flip line, associated with the reordering of the ferromagnetic component of the magnetization, and an intriguing line for field induced magnetism, also corroborated by magnetization measurements.}
\end{abstract}
 % end of PACS codes

\maketitle

\section{Introduction}
\label{intro}
In the recent years, iridium oxides have become a new playground for the study of electron correlation effects. Indeed, while extended 5$\textit{d}$ orbitals  reduce the electron-electron interaction, as compared to the 3$\textit{d}$ transition metal compounds as cuprates, strong spin orbit coupling (SOC) associated to the heavy Ir competes, together with the on-site Coulomb interaction, with electronic bandwidth to restore such correlations\cite{Kim2008}. Amongst these compounds, the Ruddlesden-Popper series, R$_{n+1}$Ir$_n$O$_{3n+1}$ where R= Sr, Ba and n = 1,2,$\infty$, has attracted much of the attention, in particular due to the structural similarities of these perovskites with the cuprates compounds. Sr$_2$IrO$_4$, where one IrO$_2$ layer alternates with an SrO layer, is structurally similar to the first discovered cuprate superconductor, (La,Ba)$_2$CuO$_4$. The physics of the latter is the one of an antiferromagnetic Mott insulator, with a magnetic interaction described within the framework of a spin-1/2 Heisenberg model. It was early proposed that the strong SOC in the iridate perovskite actually allows for an effective localized state, entangling spin and orbital degrees of freedom, with total angular momentum J$_{eff}$ = 1/2. This spin-orbital insulating state was proposed to be the analog of a Mott insulator \cite{Kim2008}.

\begin{figure}
\resizebox{0.75\columnwidth}{!}{%
  \includegraphics{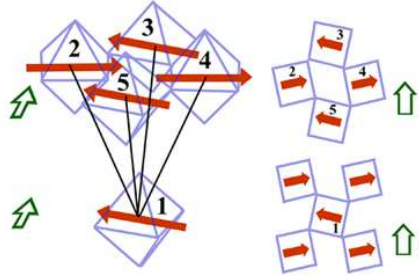}
  }
\caption{With ferromagnetic coupling of the in-plane magnetic moment, as shown (hollow arrow), transverse hopping occurs within a magnetic sublattice of equivalent Ir, 1-5-3. Antiferromagnetic coupling is obtained reversing the spins in one layer, and hopping within a sublattice occurs between inequivalent 1-2-4.}\label{sketchIr}
\end{figure}

The antiferromagnetic order in Sr$_2$IrO$_4$ is now well documented\cite{Kim2009,Ye13}. The moments (0.2 $\mu_B$/Ir) lay in-plane  and order at $T_N$ $\simeq$ 240~K. The loss of the inversion symmetry in the non cubic structure, due to a rotation of the oxygen octahedra, allows for a Dzyaloshinskii-Moriya interaction, which in turn induces a canting of the spins and a ferromagnetic component in the IrO$_2$ planes \cite{Crawford94,Jackeli09} (Fig.~\ref{sketchIr}). The net moment ($\mu$ = 0.14 $\mu_B$/Ir), which is coupled in an 'up-up-down-down' way from plane to plane in zero field, align ferromagnetically with an in-plane field $H \approx$ 0.2~T \cite{Cao98}. Recent \textit{ab initio} computations conclude that the dominant magnetic interaction is of Heisenberg type, with little effect of the geometrical factors on the exchange coupling \cite{Katukuri12}. As shown in ref.~\cite{Fujiyama12}, the absence of a critical behavior in the in-plane magnetic correlation length at $T_N$ is also in favor of a two-dimensional Heisenberg behavior with large quantum fluctuations. On the basis of such a description, it has been proposed that Sr$_2$IrO$_4$ may exhibit electronic properties similar to the ones of the cuprates, including superconductivity \cite{Wang11}.
The nature of the insulating state is however the subject of debate. First, the realization of the $J_{eff}$ = 1/2 state itself may be questioned, as it requires a perfect orbital degeneracy, which is not obtained in Sr$_2$IrO$_4$ where the octahedra are strongly elongated \cite{MorettiSala14}. The location of a metal-insulator transition, either in the paramagnetic state as for a Mott-Hubbard transition \cite{Martins11}, or coincident with the magnetic transition as for a Slater-type transition \cite{Arita12} is controversial.

As first noticed by Kini \textit{et al} \cite{Kini06}, no anomalies can apparently be detected in resistivity at $T_N$. The authors proposed that this could result from the fact that localized states shifts the Fermi level away from the band edges affected by spin polarization. A time-resolved optical study found that the metal-insulator transition takes place over a wide temperature range 0.7 $\lesssim T/T_N$ $\lesssim$ 1.4, thus accounting for the absence of any sharp anomaly in transport and thermodynamic quantities\cite{Hsieh12}. Well below $T_N$, for $T \lesssim$ 100~K, large anisotropic magnetoresistance as well as magnetodielectric effects were observed (Refs.~\cite{Chikara09},\cite{Ge11}). It was proposed that the magnetoelectric effects result from the competition of antiferromagnetic and ferromagnetic coupling, at low and high temperature respectively\cite{Chikara09}. Recent mesoscopic anisotropic magnetoresistance measurements at low temperature also proposed a coupling between the quadratic crystalline structure; the orthorhombic magnetic one, and transverse electronic transport\cite{Wang14}.

Here, we show that there is actually a small but clear signature of the magnetic transition in zero field in the transverse resistivity. We propose that this resistivity change may be evaluated considering the hopping of the localized charge. In an applied field, it is shown that magnetotransport  also tracks the magnetic phase diagram. In particular, both resistivity and magnetization measurements point toward field-induced antiferromagnetism.

% ~~~~~~~~~~~~~~~~~~~~~~~~~~~~~~~~~~~~~~~~ checked

\section{Resistivity}

\begin{figure}
\resizebox{0.75\columnwidth}{!}{%
  \includegraphics{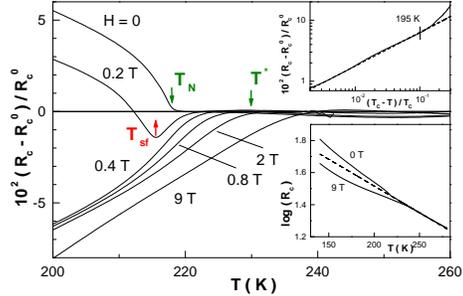}
  }
\caption{\textit{c}-axis resistance anomaly (sample 1), as obtained subtracting a high temperature linear logarithmic resistance (dashed line in lower inset). Upper inset: scaling of the resistive anomaly. The magnetic field is applied at $\approx$ 10 deg. from the \textit{c}-axis, and the in-plane field component accounts for the shift of the positive anomaly to lower temperature $T_{sf}$.}\label{Rperp}
\end{figure}

The results below where obtained from two Sr$_2$IrO$_4$ single crystals for transport measurements, with dimensions 300 x 200 x 30 $\mu m^3$ (sample 1) and 500 x 200 x 100 $\mu m^3$ (sample 2), as well as from a larger sample for squid magnetization measurements, with dimensions 1500 x 500 x 500 $\mu m^3$ (sample 3). They were grown using a self-flux technique in platinum crucibles, similar to the one in Ref.~\cite{Kim2009}. We denote \textit{a} and \textit{b} the crystal lattice vectors of the superstructure in the IrO$_2$ planes\cite{Crawford94}. Magnetization of these crystals showed an onset at $T$ = 220-240 K, which was dependent upon the applied field, as discussed below. The zero field ordering temperature, $T_N$, was respectively 217 $\pm$ 1 K for sample 1 and 2 and 220 $\pm$ 1 K for sample 3, as determined below. The typical resistivity ratio for these samples, $\rho_c(10 K)/\rho_c(300 K) = 2 \, 10^3$ was indicative of a low doping content\cite{Korneta10}. Low-resistance contacts were achieved using silver epoxy annealed at 500 °C in oxygen atmosphere.

Careful investigation of the \textit{c}-axis resistivity reveals the existence of an anomaly at a temperature close to the reported $T_N$ for the undoped material. The anomaly is actually very small (Fig.~\ref{Rperp}, lower inset), but the sharp jump in the resistivity temperature derivative in zero magnetic field unambiguously points towards a well defined phase transition, at $T_N$ = 217$\pm$ 0.5~K for sample 1 (Fig.~\ref{Rperp}). This positive MR observed at zero field becomes negative at 9 T. In the following, we analyze the evolution from the positive to the negative anomaly, and what it reveals on the coupling between resistivity and magnetic structure.

Quantitatively, the evaluation of the resistance change below $T_N$ requires subtracting some arbitrary background, as obtained from the high temperature resistivity. We have used a linear fit for the resistance logarithm at high T (dashed line in lower inset of Fig.~\ref{Rperp}). At least close to the transition temperature, little error is likely made due to the sharp transition, and a scaling of the resistivity change may be attempted, which is found to hold within a $\approx 20$~K interval (Fig.~\ref{Rperp}, upper inset). We did not find any influence of the magnetic transition on the in-plane resistivity.

\begin{figure}
\resizebox{0.75\columnwidth}{!}{%
  \includegraphics{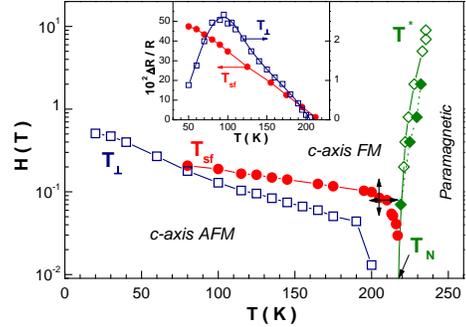}
  }
\caption{Phase diagram for the transverse resistivity, as obtained from data as in Fig.~\ref{MR} (squares and circles) and the onset in Fig.~\ref{Rperp} (diamonds) (full symbols are for field applied along \textit{a}(\textit{b}); filled ones, along \textit{c}). $T_N$ is the zero field transition temperature, as given by the data in Fig.~\ref{Rperp}. The inset displays the amplitude of the magnetoresistance at the $T_{sf}$ and $T_\bot$ lines (sample 2).}\label{phasediag}
\end{figure}

\begin{figure}
\resizebox{0.75\columnwidth}{!}{%
  \includegraphics{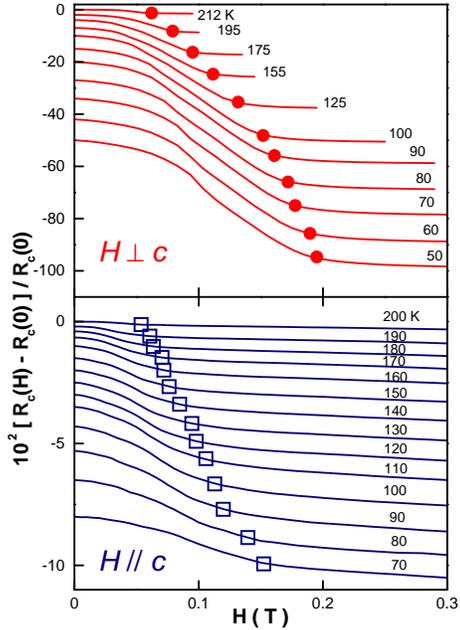}
  }
\caption{\textit{c}-axis magnetoresistance, for field along and perpendicular to the \textit{c}-axis (curves have been shifted from their $H=0$ zero value, for clarity) (sample 2).}\label{MR}
\end{figure}

\textit{Below} $T_N$, with an applied magnetic field, we identify a line for each of the \textit{a}(\textit{b})-axis and the \textit{c}-axis field orientations, denoted respectively $T_{sf}$ and $T_\bot$ (Fig.~\ref{phasediag}). $T_{sf}$ may be tracked as the shift to \textit{lower} temperature, with the magnetic field \textit{in-plane} component, of the \textit{positive} magnetoresistance anomaly (as an example, data in Fig.~\ref{Rperp} allows to assign a transition temperature $T_{sf}$ = 215.5 ~K for $H_\bot \approx $ 0.2 $\sin(10) \approx$ 0.035~T, where 10 deg. is the estimated tilt of the magnetic field from the \textit{c}-axis). $T_{sf}$ retains its sharp character at small field, but, at larger field, the anomaly is quickly washed out. However, we find that a kink in the magnetoresistance substitutes to this anomaly (Fig.~\ref{MR}), allowing to extend the definition of $T_{sf}$ to low temperature in Fig.~\ref{MR} (circles). 

\begin{figure}
\resizebox{0.75\columnwidth}{!}{%
  \includegraphics{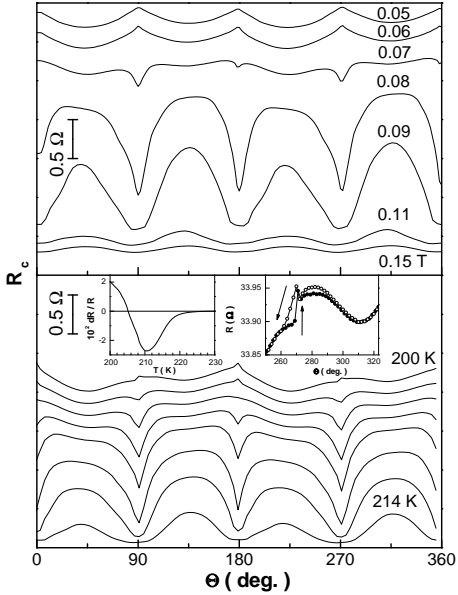}
  }
\caption{Angular dependence of the magnetoresistance along the two segments in Fig.~\ref{phasediag} (field in the \textit{a}-\textit{b} plane; sample 1). $\theta$ is the angle between the bisector of \textit{a} and \textit{b} and the magnetic field. Upper panel: $T$ = 204 K. Lower panel: $H$ = 0.07 T ($T$ varies in steps of 2 K);  left inset: resistance anomaly in zero field, obtained as in Fig.~\ref{Rperp}; right inset: occurrence of a uniaxial hysteretic feature ($T$ = 200 K).}\label{angular2}
\end{figure}

The angular dependence of the magnetoresistance along constant $T$ and $H$ intervals crossing $T_{sf}$ (Fig.~\ref{angular2}) displays two features: i) a large angular susceptibility of the transverse magnetoresistance develops at the crossing of the line along the \textit{a} and \textit{b} directions; ii) just below this line (at 200 K in Fig.~\ref{angular2}), there is a two-fold periodicity (we have checked, deliberately tilting the crystal and observing no qualitative change in the $R_c(\theta)$ behavior, that this cannot be due to the sample misalignment), and the peaks in angular susceptibility is hysteretic. The magnitude of the resistance change at $T_{sf}$ and $T_\bot$ is found to depend on temperature in strikingly different manners: while the magnetoresistance at $T_{sf}$ increases monotonously with decreasing temperature, it peaks at $T \approx$ 100 K for $T_\bot$. The resistivity drop at $T_\bot$ is also one order of magnitude smaller than at $T_{sf}$ (Fig.~\ref{phasediag}, inset).

\begin{figure}
\resizebox{0.75\columnwidth}{!}{%
  \includegraphics{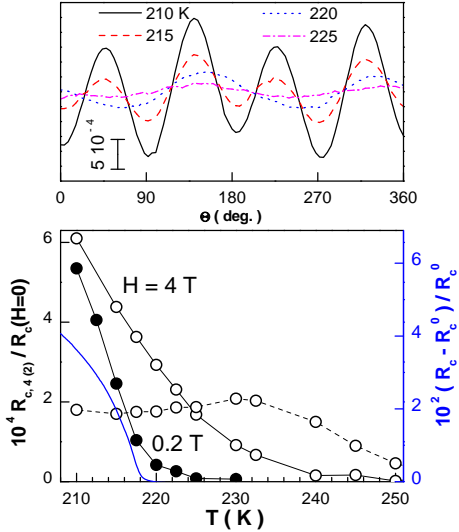}
  }
\caption{Upper panel: angular dependence of the magnetoresistance (in-plane field $H$ = 0.2 T). Lower panel: Symbols: four-fold and two-fold (dotted line) components of the angular \textit{c}-axis magnetoresistance (left scale). Full line: zero field \textit{c}-axis resistance anomaly, as in Fig. \ref{Rperp} (sample 1).}\label{angular}
\end{figure}

\textit{Above} $T_N$, a \textit{negative} contribution to the magnetoresistance shows up, the onset of which is weakly dependent upon the field orientation. This allows to track $T^*(H)$ (Fig.~\ref{phasediag}, open symbols), as will be seen below. The positive slope for $T^*(H)$ implies that the transition temperature obtained from high field studies (as in magnetometry) must be overestimated, as will be seen below. Data in Fig.~\ref{Rperp} suggest that a maximum shift $T^* - T_N\approx$ 20~K is obtained for $H \approx$ 7~T. As noticed above, the procedure to evaluate the resistance change at the transition is quite arbitrary, which may be a problem in the case of a smooth variation as observed at $T^*$. Angular dependent magnetoresistance, however, confirm the general trend for this line. Indeed, rotating a large magnetic field around the \textit{c}-axis reveals the existence of a four-fold contribution to the (negative) magnetoresistance (Fig.~\ref{angular}). The angular-dependent contribution is typically only a few percent of the total magnetoresistance in Fig.~\ref{MR}, and is found maximal along the \textit{a} and \textit{b} axis of the crystal. The temperature dependence of the four-fold component extracted in Fig.~\ref{angular} confirms, with no need for a high-temperature background fit, the existence of an onset at $T^*$. A two-fold angular component is also present, which vanishes with increasing temperature simultaneously with the four-fold component. However, while the four-fold component temperature dependence appears similar to the one of some order parameter (as for the zero field anomaly), the two-fold component saturates with decreasing temperature.

Applying a magnetic field either along\textit{a}(\textit{b})-axis or along the \textit{c}-axis requires a cautious consideration of the possible misalignment effects. In a first approximation, assuming a quasi-two-dimensional behavior with the easy axis in the \textit{ab} plane, the perpendicular magnetoresistance may be expected to be strongly affected by a tilt from the \textit{c}-axis direction, while the in-plane one is expected weakly sensitive to misalignment. So, we have checked that the line for the \textit{c}-axis field orientation, $T_\bot$, truly originates from the perpendicular field component: sample 2, measured with a magnetic field at 2.5 deg. and 5 deg. from the \textit{c}-axis, showed that the resistivity drop at this line cannot be scaled to any longitudinal magnetoresistance drop, and that this line is independent of the small tilt angle value. The large negative magnetoresistance at $T_{sf}(H)$ that we observe at low temperature is similar to the one reported earlier by Ge \textit{et al} below $T$ = 100 K (Ref.~\cite{Ge11}). The magnetoresistance resistance anomaly at large transverse field in Ref.~\cite{Ge11} ($H \approx$ 3 T at $T$ = 50 K) is attributed here to the replica of the $T_{sf}$ line, due to misalignment, which could also be observed here for a large tilt.

\section{Discussion}

We first comment on the $T_{sf}$ line. We attribute this line to the spin-flip mechanism, which re-orients the \textit{c}-axis 'up-up-down-down' arrangement for the in-plane ferromagnetic component to a ferromagnetic one (Fig.~\ref{sketchIr}. Two contributions should actually be distinguished: one due to the change in the transfer integrals and linked to the orientation of the IrO$_6$ octahedra, as proposed in Ref.~\cite{Ge11}, the other one as a pure spin configuration effect where, depending on the spins orientation, hopping of the localized charge within one sublattice occurs between slightly different relative spin configurations. Here, we examine whether the latter mechanism could contribute in the present case.

The problem of the observation of a magnetoresistance -- as large as in the present case -- associated to the presence of a weak ferromagnetic state induced by the magnetic field was already encountered in the case of cuprates\cite{Thio88,Thio90}, and received a quantitative interpretation in the case of La$_2$CuO$_4$\cite{Shekhtman94}. In La$_2$CuO$_4$, in the orthorhombic phase, the CuO$_6$ octahedra are tilted from the CuO planes by $\alpha \approx$ 3 deg. (inducing a ferromagnetic component perpendicular to the planes). This allows an antisymmetric superexchange term in the spin Hamiltonian, which would otherwise be zero due to symmetry in the tetragonal phase\cite{Thio88}. In Sr$_2$IrO$_4$, IrO$_6$ octahedra are tilted in the IrO planes by $\alpha \approx$ 11 deg. (inducing a ferromagnetic component in the planes) and the compound is tetragonal. This also destroys the inversion center which exists midway between the Ir atoms, and allows for a non zero antisymmetric superexchange term\cite{Crawford94} (Fig.~\ref{sketchIr}). Equivalently, due to the tilt of the IrO$_6$ octahedra, the transfer integral between one Ir atom and its four nearest neighbors in the next adjacent plane are unequal. As a consequence, interchanging magnetic sublattices, as could be induced by a spin flip, may strongly influence the transverse conductivity in this case also.

We observe that the transverse resistivity can be fitted with the conventional expression for three-dimensional VRH, $\rho \propto \exp{(T_0/T)}^{1/4}$, using $T_0 = 6 \, 10^5$~K (where $T_0 \propto \lambda^{-3}$, and $\lambda$ is the transverse localization length -- see Ref.~\cite{Shekhtman94} and Refs therein). This yields for the ratio of the hopping length to the localization radius  $l/\lambda \approx (T_0/T)^{1/4} \approx 7$, indicating that charge hopping from impurity centers is controlled by the pure material transfer integrals between sites, as was proposed in the case of La$_2$CuO$_4$\cite{Shekhtman94}. Using $\lambda \propto t^{1/2}$, where $t$ is some effective transfer integral between planes, we expect in this case a relative change $\delta\rho/\rho \approx -\frac{3}{8}(T_0/T)^{1/4} \delta t/t$. According to Ref.~\cite{Shekhtman94}, a flip of the spins required to align ferromagnetic moments in the plane is associated to a change $\delta t/t \approx \frac{1}{2} \, (J_\perp/J_\parallel) (\kappa_\parallel/\kappa_{\perp})^4 (m_{\perp}/m_\parallel)^2$, where $m_{\parallel(\perp)}$ is the in-plane (transverse) effective mass, $J_{\parallel(\perp)}$ is the in-plane (transverse) exchange coupling, and $\kappa_{\parallel(\perp)}$ is the corresponding reciprocal lattice constant. We have $\kappa_\parallel/\kappa_{\perp} = 1.8 $, $J_\perp/J_\parallel \approx 10^{-5} $ (a value comparable to that for La$_2$CuO$_4$, Ref.~\cite{Fujiyama12}) and $m_{\perp}/m_\parallel > 20$ (this is evaluated by the ratio of the bandwidth for J=1/2 along $\Gamma$X and NC\cite{Arita12}). This yields $\delta \rho/\rho = \frac{1}{2}\, \delta t/t > 6\, 10^{-2}$. Though this is only a rough estimate and magnetic configurations for both cases are different, this illustrates that an effect comparable to the one that we observe may be expected from the spin contribution alone. We expect this contribution to be significant at the $T_{sf}(H)$ line, where there is a field-induced ferromagnetic moment\cite{Ge11}.

The observations in Fig.~\ref{angular2} then receive a straightforward interpretation: at the spin-flip transition, there is a divergence of the magnetic susceptibility, which may be evidenced as one crosses the surface which marks this transition (in a temperature-orientation space -- the $T_{sf}$ line being the intersection of this surface with a constant orientation surface). Ferromagnetic domains being linked to the possibility to order magnetism from plane to plane, it is natural to expect their hysteretic signature to show up below this transition, as well as some potential \textit{a}/\textit{b} unbalance due to inequivalent domains.

Within this hypothesis that the resistivity change results from the larger transverse transfer integral associated to magnetic ordering, we may tentatively relate the observed scaling of the resistivity to a critical exponent. We expect the resistance change to be proportional to the phase transition order parameter associated to inter-plane spin ordering, $M$ (this may be assumed in the framework of a two-fluid model, for which there is an amount $n \propto M$ of ordered moments associated to a larger transfer integral). As a result, the scaling exponent for the resistivity is identical to the conventional exponent $\beta$ for the order parameter. The value obtained, $\beta \simeq 0.55$, is close to the one for a mean field type transition ($\beta = 0.5$). In Ref.~\cite{Fujiyama12}, the transverse fluctuation correlation length above $T_N$ yielded a critical exponent $\nu = 0.75 \pm 0.05$. This value is far off the mean field value ($\nu = 0.5$). This discrepancy could sign the limit of the present analysis for the resistivity scaling, made within a simple static picture. Also, the scaling in Ref.~\cite{Fujiyama12} relates to long range correlations ($\simeq 3-20$ \textit{c}), while we expect resistivity to be essentially driven by magnetic correlations at the scale of the inter-plane distance. This scenario might also provide some hints to the understanding of magnetotransport in Sr$_3$Ir$_2$O$_7$ (327). The magnetic structure of this compound is indeed very different from the one of Sr$_2$IrO$_4$, with an out-of-plane collinear antiferromagnetic ordering for the latter\cite{Fujiyama12b}. In-plane and out-of-plane resistivity also strongly increase at the magnetic transition, for 327, in contrast to 214 (Ref.~\cite{Cao02}). It is then possible that hopping between antiferromagnetically coupled spins in the ordered phase of 327 dictates a much larger resistivity change than the one due to the slightly inequivalent spins when \textit{c}-axis correlation is lost in 214.
 
We now comment on the line for transverse magnetic field, $T_{\bot}$. The downturn of the magnetoresistance effect amplitude below $T \approx$ 100 K appears specific to this line. We have checked, tilting sample 2 at 2.5 deg. and 5 deg. from the \textit{c}-axis, that the small tilt angle has no role in this non-monotonous behavior. This observation strongly evokes previous ones by Chikara \textit{et al} of a giant magnetoelectric effect at a comparable temperature\cite{Chikara09}. The authors interpreted this observation as the result of the competition between a ferromagnetic exchange coupling and an antiferromagnetic one, the latter becoming dominant below $T \approx$ 100 K, as a result of the progressive decrease of the Ir-O-Ir angle with decreasing temperature (i.e. a larger spin canting). The downturn of the magnetoresistance effect amplitude is, however, observed here independent of the ordering of the ferromagnetic component along the \textit{c}-axis by the applied field, and the related anomaly at $T_{sf}$ displays no accident which could sign a change in the \textit{c}-axis coupling (Fig.~\ref{phasediag}, inset). The magnitude of the magnetoresistance anomaly at $T_{\bot}$ -- about one order of magnitude smaller than at $T_{sf}$ -- is also in favor of marginal spin reorientation, as compared to the one involved in a spin-flip transition. The out-of-plane tilt of the spins for such a modest transverse magnetic field as the one observed along the $T_{\bot}$ line is expected to be quite small (of the order of $H g \mu_B S/6k_B T_N$), as it is well below the paramagnetic critical field. The observation of a \textit{c}-axis resistivity change suggests, however, that a modification in the \textit{c}-axis spin coupling may come with this tiny spin reorientation. This is plausible, as the transverse exchange spin coupling is $J_{\bot} \approx \mu H$, where $H \approx 0.1$ T. While the exact mechanism coupling a transverse magnetic field to the spin configuration is not known, we could put an upper bound to the structural changes that have been evoked to account for such magnetotransport effects\cite{Ge11}. Preliminary measurements indeed indicate that a 0.6 T transverse magnetic field at 150 K (i.e. crossing the $T_{\bot}$ line), does not modify the \textit{c}-axis parameter by more than $dc/c$ = $2 \, 10^{-5}$. So, a pure spin effect could be at play here also.

\begin{figure}
\resizebox{0.75\columnwidth}{!}{%
  \includegraphics{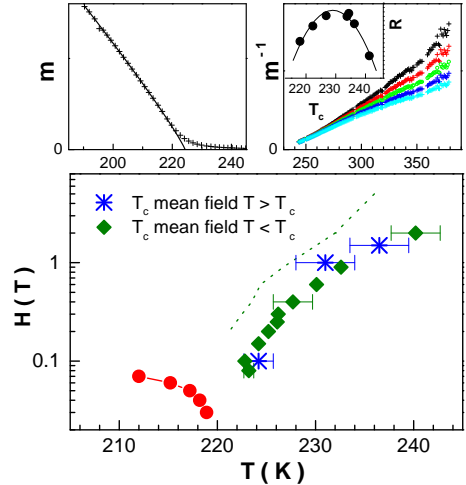}
  }
\caption{Upper left panel: determination of $T_c$ from a fit of the magnetization for $T < T_c$ to a power law ($H$ = 0.1 T). Upper right panel: determination of $T_c$ from a fit of $\chi^{-1}(T)$ to a line for $T > T_c$, maximizing the fit regression factor ($R$) by tuning the background signal, optimum at the middle curve displayed in the panel ($H$ = 1 T). Lower panel: resulting phase diagram (field applied along \textit{a}(\textit{b}), sample 3). The dotted line marks an arbitrary deviation from the power law fit of the magnetization below $T_c$, and full circles is spin-flip as obtained from the magnetization jump.}\label{mag}
\end{figure}

Finally we consider the unusual behavior for $T^*(H)$, showing field induced magnetism above $T_N$. The signature in the transverse resistivity, occurring in the vicinity of the $T^*(H)$ line, signs both the occurrence of the in-plane quadratic ordering (as evidenced by the four-fold symmetry) and of the \textit{c}-axis ordering (which is necessary to account for a transverse magnetotransport effect). The transition being between the paramagnetic state and the field-induced \textit{c}-axis aligned state (it is likely that a small in-plane orients the ferromagnetic component in the transverse configuration also), it can be of a different sign than the zero-field transition between the paramagnet and the \textit{c}-axis 'up-up-down-down' configuration.

Field induced magnetism does show up in the magnetization data also. It was shown in early studies that the Curie-Weiss temperature, $T_c$, characterizing the hidden weak ferromagnetism below $T_N$ of this compound, may be accurately determined both from the onset of magnetization \textit{below} $T_N$, and in the paramagnetic regime \textit{above} $T_N$ ($T_c$ = 234.6 K in Ref.~\cite{Cao98}). While it is very often made no distinction between the Neel temperature and the Curie-Weiss temperature, as the ferromagnetic component arises from the 3D magnetic ordering, we will keep here this distinction, since the analysis of the magnetization data provides $T_c$ -- strictly speaking. The critical behavior underlying the Curie-Weiss behavior is the one of the \textit{c}-axis magnetic correlations\cite{Fujiyama12}.

\textit{Above} $T_c$, we performed Curie-Weiss fits of the magnetization data not too close to $T_c$ (where critical fluctuations are dominant) and for temperatures less than $\approx 2\, T_c$ (above which the Bethe first approximation shows that the actual $T_c$ is lower than expected from the fit). As the fitting parameter $T_c$ sensitively depends on the subtraction of a background signal from the sample holder, we introduced this background as an extra parameter, which was determined as the one minimizing the deviation of the inverse susceptibility data from a linear behavior (Fig.~\ref{mag}, upper right panel). \textit{Below} $T_c$, we have fitted the magnetization, for fields larger than the spin-flip one, to a power law (mean field analysis of the ferromagnetic order parameter, Fig.~\ref{mag}, upper left panel). Both transition temperatures agree well, which validates the analysis. $T_c(H)$ also reproduces the behavior for $T^*$, increasing with the magnetic field. So, three experimental signatures point towards some enhancement of the magnetic correlations with the applied field, and the emergence of the $T^*$ and $T_c$ lines from the zero field Neel point $T_N$ questions a possible increase of the latter with magnetic field.

A cause for this re-entrant behavior could be the presence of a competing magnetic order, but the examples we are aware of require geometric frustration of the AF order, which is not present here. As evidenced from the 3D ordering temperature of a 2D Heisenberg AF magnet with weak transverse coupling (see e.g. Ref.~\cite{Thio88}), $k\, T_N = (\xi_{2D} / a)^2 J_{\bot}$, where $\xi_{2D}$ is the in-plane magnetic correlation length and $a$ is the magnetic lattice spacing, magnetic ordering may be promoted both by an enhancement of the 2D AF magnetic correlations, and by an increase of the transverse coupling. The effect of an effective staggered field, obtained in the presence of the Dzyaloshinskii-Moriya interaction and an applied field\cite{Hamad2005,Oshikawa1997,Moskvin2007,Kagawa2008}, puts the first of these two mechanisms at play. This was invoked in the case of La$_2$CuO$_4$, to account for $^{17}O$ Knight shift anomalies in the paramagnetic state\cite{Moskvin2007}. Interestingly, there should be a four-fold component for this effect (maximum along the Ir-O-Ir bond), as observed here. The larger spin-orbit coupling in the present case would contribute to a larger effect. Another mechanism, where the magnetic field aligns the ferromagnetic moment of short range fluctuating ordered domains, is also conceivable. We believe it belongs to the second kind of mechanism, introducing an effective transverse coupling in the presence of the magnetic field. It is however difficult to explain in this case that the effect on the resistivity is maximum when the field is aligned at 45 deg. from the ordered ferromagnetic direction (Fig.~\ref{angular}). To further discriminate in favor of one of these two mechanisms, and determine whether it just contributes to enhance magnetic correlations or promote a true magnetic order (and thus shifts $T_N$), requires further work, such as a direct observation of these correlations, beyond the scope of this paper.\newline

In conclusion, our data clearly contradicts the general belief that there would be no influence of the transition to the ordered magnetic state on the transport properties. We have shown that the transverse resistivity actually allows to track the entire magnetic phase diagram of our compound, bearing the signatures of the spin-flip transition, of a transverse magnetic field one, and of field-induced antiferromagnetism.  We propose that magnetotransport is influenced strongly by direct spin reorientation effects, in addition to possible bond reorientation ones, and that this influence can be evaluated considering hopping of the localized charge.\newline

L.F. performed the experiments and wrote the paper, with inputs from all co-authors, who also provided samples. V.B. thanks the group of I.R. Fisher (Stanford university) for initial help with sample growth.

%
% For one-column wide figures use
%\begin{figure}
% Use the relevant command for your figure-insertion program
% to insert the figure file.
% For example, with the option graphics use
%\resizebox{0.75\textwidth}{!}{%
%  \includegraphics{leer.eps}
%}
% If not, use
%\vspace{5cm}       % Give the correct figure height in cm
%\caption{Please write your figure caption here}
%\label{fig:1}       % Give a unique label
%\end{figure}
%
% For two-column wide figures use
%\begin{figure*}
% Use the relevant command for your figure-insertion program
% to insert the figure file. See example above.
% If not, use
%\vspace*{5cm}       % Give the correct figure height in cm
%\caption{Please write your figure caption here}
%\label{fig:2}       % Give a unique label
%\end{figure*}
%
%
% BibTeX users please use
% \bibliographystyle{}
% \bibliography{}
%
% Non-BibTeX users please use

\end{document}